\documentclass[lettersize,journal]{IEEEtran}
\usepackage{amsmath,amsfonts}
\usepackage{algorithmic}
\usepackage{array}
\usepackage{amsmath}
\usepackage{amssymb}
\usepackage[caption=false,font=normalsize,labelfont=sf,textfont=sf]{subfig}
\usepackage{textcomp}
\usepackage{stfloats}
\usepackage{url}
\usepackage{verbatim}
\usepackage{graphicx}
\usepackage{graphicx}
\usepackage{multirow}
\usepackage{graphicx}
\usepackage{cite}
\usepackage{orcidlink}
\usepackage{hyperref}
\usepackage[noabbrev, capitalise]{cleveref}

\def\BibTeX{{\rm B\kern-.05em{\sc i\kern-.025em b}\kern-.08em
    T\kern-.1667em\lower.7ex\hbox{E}\kern-.125emX}}
\usepackage{balance}
\begin{document}
\title{Cyber Mobility Mirror for Enabling Cooperative Driving Automation in Mixed Traffic: \\A Co-Simulation Platform}
\author{
Zhengwei Bai$^{\orcidlink{0000-0002-4867-021X}}$,~\IEEEmembership{Student Member, IEEE},
Guoyuan Wu$^{\orcidlink{0000-0001-6707-6366}}$,~\IEEEmembership{Senior Member, IEEE},
Xuewei Qi,~\IEEEmembership{Member, IEEE},
Yongkang Liu,
Kentaro Oguchi,
Matthew J. Barth$^{\orcidlink{0000-0002-4735-5859}}$,~\IEEEmembership{Fellow, IEEE}


\thanks{Zhengwei Bai, Guoyuan Wu, and Matthew J. Barth are with the Department of Electrical and Computer Engineering, University of California at Riverside, Riverside, CA 92507 USA (e-mail: zbai012@ucr.edu).}

\thanks{Xuewei Qi, Yongkang Liu, and Kentaro Oguchi are with the Toyota North America R\&D Labs, Mountain View, CA 94043, USA.}

}

\markboth{IEEE Intelligent Transportation Systems Magazine}%
{How to Use the IEEEtran \LaTeX \ Templates}

\maketitle

\begin{abstract}
Endowed with automation and connectivity, Connected and Automated Vehicles are meant to be a revolutionary promoter for Cooperative Driving Automation. Nevertheless, CAVs need high-fidelity perception information on their surroundings, which is available but costly to collect from various onboard sensors as well as vehicle-to-everything (V2X) communications. Therefore, authentic perception information based on high-fidelity sensors via a cost-effective platform is crucial for enabling CDA-related research, e.g., cooperative decision-making or control. Most state-of-the-art traffic simulation studies for CAVs rely on the situation-awareness information by directly calling on intrinsic attributes of the objects, which impedes the reliability and fidelity of the assessment of CDA algorithms. In this study, a \textit{Cyber Mobility Mirror (CMM)} Co-Simulation Platform is designed for enabling CDA by providing authentic perception information. The \textit{CMM} Co-Simulation Platform can 
emulate the real world with a high-fidelity sensor perception system and a cyber world with a real-time rebuilding system acting as a ``\textit{Mirror}" of the real-world environment. Concretely, the real-world simulator is mainly in charge of simulating the traffic environment, sensors, as well as the authentic perception process. The mirror-world simulator is responsible for rebuilding objects and providing their information as intrinsic attributes of the simulator to support the development and evaluation of CDA algorithms. To illustrate the functionality of the proposed co-simulation platform, a roadside LiDAR-based vehicle perception system for enabling CDA is prototyped as a study case. Specific traffic environments and CDA tasks are designed for experiments whose results are demonstrated and analyzed to show the performance of the platform.

\end{abstract}


\section{Introduction}
\IEEEPARstart{W}{ith} rapid development of the economy and society, the field of transportation is facing several major challenges caused by drastically increased traffic demands, such as improving traffic safety, mitigating traffic congestion, and reducing mobile source emissions. Cooperative Driving Automation (CDA) enabled by Connected and Automated Vehicles (CAVs) is regarded as a promising solution to the aforementioned challenges \cite{fagnant2015preparing}. In the past few decades, several projects have been conducted to explore the potential of CDA. The California PATH program \cite{misener2006path} demonstrated the improvement of traffic throughput by an automated platoon utilizing connectivity. The European DRIVE C2X project \cite{stahlmann2011starting} assessed the cooperative system by large-scale field operational tests of various connected vehicle applications. Fujitsu has launched a \textit{Digital Twin} platform for supporting mobility by connectivity and artificial intelligence~\cite{2020Fuji}. These projects have demonstrated CDA to be a transformative path toward the next-generation transportation system, which is enabled by ubiquitous perception, seamless communication, and advanced artificial intelligence technologies.
\begin{figure}[!t]
    \centering
    \includegraphics[width=0.5\textwidth]{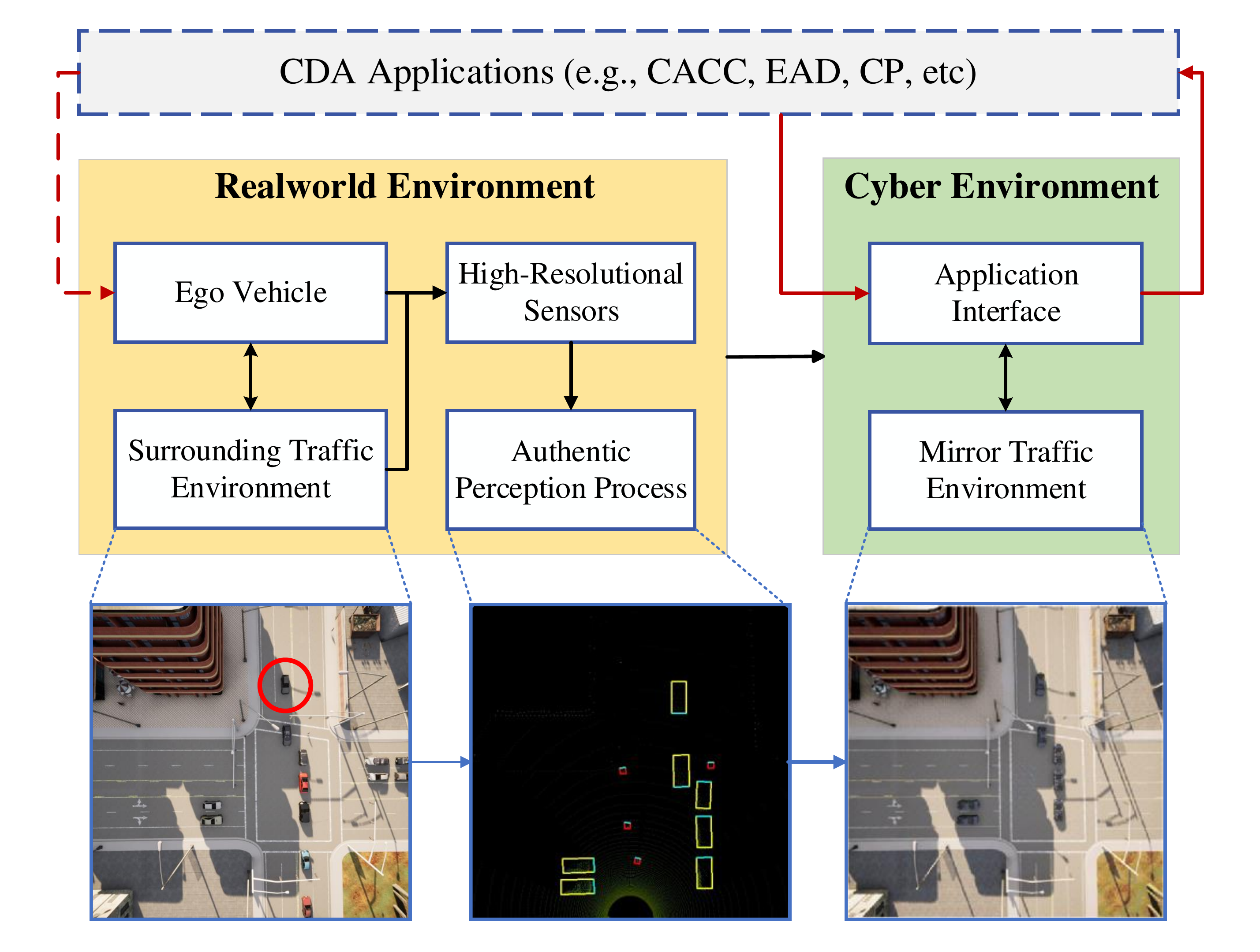}
    \caption{The systematic architecture for the proposed CMM co-simulation platform and the workflow for supporting CDA. Three sample figures, including real traffic environment (left), perception results visualization (mid), and cyber environment (right), are shown for illustrating the main goal of the platform.}
    \label{fig:systematic}
\end{figure}

Given the fact that the cost of large-scale real-world deployment is prohibitive, it is imperative to design and assess CDA systems based on simulation. For instance, as one key component of a CDA system, connected and automated vehicle (CAV) technologies heavily rely on simulation to comprehensively assess their performance in terms of safety, efficiency, and environmental sustainability. Therefore, simulation platforms for CDA are of great significance and their development receives much attention. 
    
Additionally, to enable CDA, accurate perception information can lay a solid foundation, which requires inputs from different types of high-fidelity sensors, such as radar, camera, and LiDAR. Direct implementation of these sensors to perceive the real-world environment may be costly or time-consuming, and in some cases, restricted by application scenarios. Thus, simulation platforms with high-fidelity sensor modeling and perception capability would provide a cost-effective alternative solution to CDA-related research.

Many existing traffic simulators have been developed to test various aspects of CDA. For instance, CARLA \cite{dosovitskiy2017carla} and SVL \cite{rong2020lgsvl} are designed for modeling autonomous vehicles (AVs), while SUMO \cite{lopez2018microscopic} targets microscopic traffic flows. From the perspective of sensing fidelity, most existing studies directly use intrinsic attributes of target objects, without considering the potential imperfection of perception for CDA models. This considerably limits the transferability and reliability of the real-world deployment of these CDA models. To the best of our knowledge, this paper is the first attempt to integrate the entire deep learning-based perception pipeline into the simulator to create a cyber mobility mirror (CMM) system, in which simulated traffic objects are authentically perceived and reconstructed with 3D representation. The high-level CDA model can leverage such perceived information from the module output in a CMM system rather than intrinsic attributes of target objects, to improve model fidelity and validate the system with more confidence.

The systematic structure of CMM based on a co-simulation platform is shown in Figure~\ref{fig:systematic}, where one simulator is designed to model the real-world traffic environment and the authentic perception pipeline, while the other simulator is used to work as a mobility mirror, i.e., reconstructing the perceived objects and presenting them. The output interface provides readily retrieved post-perception data for CDA applications.

The main contributions of the paper can be summarized as follows: 1) This paper proposes a cyber mobility mirror (CMM) architecture for enabling CDA research and development; 2) A prototype CMM system is designed and developed using a CARLA-based co-simulation platform; 3) A CARLA-based 3D object detection training dataset is presented; and 4) Based on the CMM co-simulation platform, a case study is conducted for roadside-LiDAR based vehicle detection to demonstrate the functionality and necessity of the proposed co-simulation platform for enabling CDA algorithm development and validation.
    
The rest of this paper is organized as follows. A brief background about traffic simulation and object perception is given in Section~\ref{background}. In Section~\ref{structure}, we first introduce the concept of CMM and then describe the design and development of the prototype system based on co-simulation. In Section~\ref{case study}, we present a case study for detecting and reconstructing vehicles based on roadside LiDAR sensing and deep learning methods, followed by the conclusion and discussion in Section~\ref{conclusion}.

\section{Background}
\label{background}
Simulation plays a crucial role in enabling cooperative driving automation, such as assessment of CAV cooperative perception algorithms and decision-making/control models~\cite{bai2022pillargrid, bai2019deep}. High-fidelity simulated sensor information lays a solid foundation for these high-level CDA algorithms and models. In this section, we briefly review the background information for simulators enabling CDA and object detection.

\subsection{Simulators Enabling Cooperative Driving Automation}
\subsubsection{Microscopic Traffic Simulators}
To model the evolution of traffic states based on traffic dynamics and interactions between traffic objects, microscopic traffic simulators have been developed for decades and greatly stimulated the development of intelligent transportation systems \cite{barcelo2010fundamentals}. These simulators mainly consist of three components: 1) transportation network defining road topology; 2) traffic generator creating traffic flows with certain demand distributions; and 3) microscopic traffic flow control strategies, including traffic signal management, vehicle driving behaviors, and moving strategies for pedestrians. 
    
Several simulation platforms are of great popularity in CDA research, such as VISSIM \cite{fellendorf2010microscopic}, Aimsun \cite{barcelo2005dynamic}, and SUMO (Simulation of Urban MObility) \cite{behrisch2011sumo}. Specifically, VISSIM and Aimsun are widely used in dealing with multi-modal traffic flow simulation due to their capabilities of providing fundamental 3D preview and statistical simulation results. SUMO is an open-source, highly portable, microscopic, and continuous traffic simulation package designed to handle large networks \cite{lopez2018microscopic}. Additionally, high compatibility to connect and interact with different kinds of external simulators, e.g., OMNeT++ \cite{varga2010omnet++}, CARLA, etc., is one of the key features of SUMO.
These microscopic traffic simulators mainly focus on general assessment for traffic dynamic performance at the network level under different traffic scenarios. Nevertheless, the design and assessment of CDA-based cooperative perception, decision-making, or control models highly rely on the fidelity of sensor data, which is a major challenge for these conventional simulators. In recent years, simulators that are capable of modeling high-fidelity sensors gain more and more interests, which are introduced in the following section.
    
\subsubsection{Autonomous Driving Simulators}
With the development of CDA, especially autonomous driving technologies (ADTs) \cite{kiran2021deep}, the requirement for high-fidelity sensors in simulators has gained more and more attention. In recent years, several autonomous driving simulators quipped with high-fidelity sensors have been developed based on game engines, such as Unity \cite{juliani2018unity} and Unreal Engine \cite{karis2013real}. For instance, AirSim \cite{shah2018airsim}, SVL, and CARLA have the capability to offer physically and visually realistic simulations for autonomous vehicle technologies (AVTs) as well as CDA systems. Specifically, AirSim includes a physics engine that can operate at a high frequency for real-time hardware-in-the-loop (HIL) simulations with support for popular protocols, such as MavLink \cite{koubaa2019micro}. SVL is a high-fidelity simulator for AVTs, which provides end-to-end and full-stack simulation that is ready to be hooked up with several open-source autonomous driving stacks, such as Autoware \cite{kato2018autoware} and Apollo \cite{graf2008apollo}. CARLA, an open-source simulator for autonomous driving, supports flexible specifications of sensor suites and environmental conditions. In addition to open-source codes and protocols, CARLA provides open digital assets (e.g., urban layouts, buildings, and vehicles) that can be used in a friendly manner for researchers. 
    
These simulators have been developed from the ground up to support the development, training, and validation of AVTs, enabling the development of CDA. They have the capacity to assess the CDA system in a cost-effective manner as well as to provide high-fidelity sensing information.
    
Although having these existing simulators, researchers still get struggled with the imperative assumption that perception data (e.g., location, velocity, rotation, etc.) is collected directly from intrinsic attributes of simulation engines, when they develop and evaluate their high-level CAV functions, such as decision-making or control methods for CDA. Therefore, to develop a generic platform that can not only support physically and visually realistic simulation but also provide perception data based on high-fidelity sensor information is still a research gap for enabling CDA system research and development.

\subsection{Object Detection in Traffic Scenes}
Object detection plays a crucial and fundamental role in enabling CDA, and it can be roughly divided into two major types: 1) traditional model-based algorithms, and 2) data-driven methods based on deep neural networks to extract hidden features from input signals and then generate detected results~\cite{bai2022survey}. The following section will briefly introduce several state-of-the-art methods for each type and high-resolution sensors, e.g., camera and LiDAR, are mainly focused on.
    
\subsubsection{Model-based Methods}
At the early stage of traffic detection, sensors with low-computational power are widely used, such as loop detectors, radar, ultrasonic, etc. Although most of them are still implemented in contemporary transportation systems, they suffer from different kinds of innate problems, such as detecting uncertainties, traffic disruption at installation, and high maintenance costs~\cite{guerrero2018sensor}. With the development of computer vision (CV) technology and the improvement of computational power, camera-based traffic object detection has been widely developed. Aslani and Mahdavi-Nasab \cite{aslani2013optical} tried to gather useful information from stationary cameras for detecting moving objects in digital videos based on optical flow estimation. In situations where limited memory and computing resources are available, Lee et al. \cite{lee2015genetic} presented a moving object detection method for real-time traffic surveillance applications based on a genetic algorithm.
    
Recently, LiDAR sensors is increasingly implemented for traffic object detection tasks due to their advantage of having higher tolerance of lighting conditions and accuracy of relative distance. Regarding traditional methods to deal with the 3D point cloud, one popular workflow is 1) background filtering; 2) traffic object clustering; and 3) object classification \cite{wu2020automatic}. Additionally, the LiDAR point cloud can also be used to identify lane markings \cite{wu2018automatic, wu2020automatic2}. Although some traditional methods have been applied to the 3D point cloud, the greater potential of LiDAR data should be tapped by data-driven methods (e.g., deep learning) which are introduced next.
    
\subsubsection{Data-driven Methods}
The development of deep neural networks (DNNs) has significantly improved the possibility of dealing with large-scale data, such as high-fidelity images or 3D point clouds. With the great success of deep learning in the image recognition area \cite{he2015spatial, bochkovskiy2020yolov4}, many DNN-based models have been implemented in object detection for traffic scenarios using cameras or LiDAR sensors. Chabot et al. \cite{chabot2017deep} presented an approach called \textit{Deep MANTA}, for multi-task vehicle analysis based on the monocular image. In terms of different lighting conditions, Che-Tsung Lin developed a nighttime vehicle detection method based on image style transfer \cite{lin2020gan}. Chen et al. \cite{chen2019smaller} proposed a shallow model named \textit{Concatenated Feature Pyramid Network} (CFPN) to detect smaller objects in traffic flow from fish-eye camera images.
    
Additionally, 3D-LiDAR is also getting more popular in traffic object detection~\cite{bai2022cyber}. For instance, Asvadi et al. \cite{asvadi2018multimodal} presented an algorithm named \textit{DepthCN} which used deep convolutional neural networks (CNNs) for vehicle detection. Considering the real-time requirement for autonomous driving applications, Zeng et al. \cite{zeng2018rt3d} proposed a real-time 3D vehicle detection method by utilizing pre-RoI-pooling convolution and pose-sensitive feature maps. Simon et al. \cite{simony2018complex} proposed an Euler-Region-Proposal for real-time 3D object detection with point clouds, called \textit{ComplexYolo}, which is capable of generating rotated bounding boxes for 3D objects. For CDA applications, traffic object detection needs meticulous consideration. Thus, in our CMM co-simulation framework, the ComplexYolo model is adapted with customized improvements for real-time vehicle and pedestrian detection.

\section{Platform Structure and Design}
\label{structure}
This section will describe the concept of cyber mobility mirror (CMM) in detail and the system architecture of the co-simulation platform based on CARLA. Specifically, the design and development of  the real-world simulator, the mirror simulator, the data communication module, and the authentic perception module will be introduced.

\subsection{CMM Based Co-Simulation Architecture}
\begin{figure*}[!ht]
    \centering
    \includegraphics[width=0.8\textwidth]{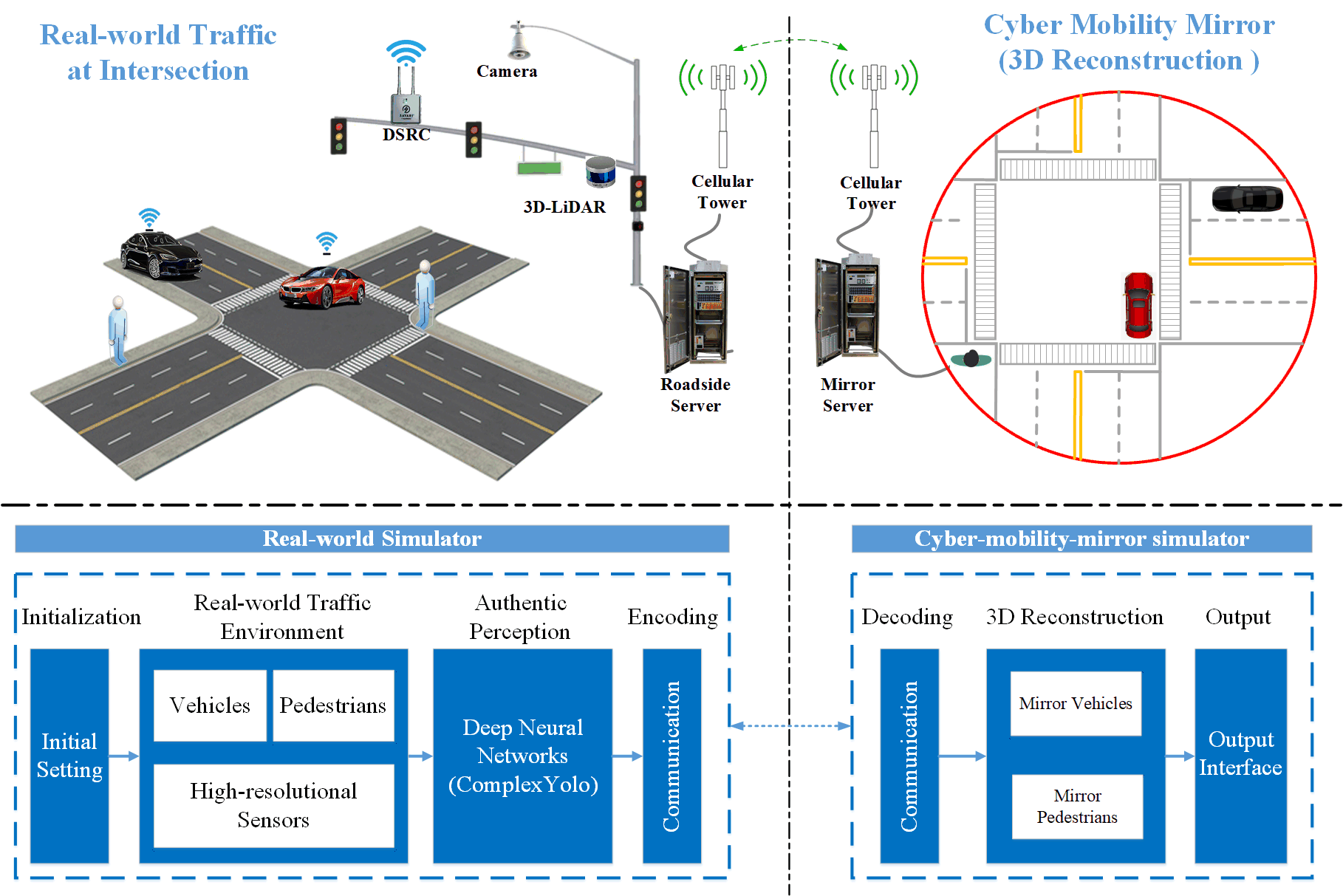}
    \caption{The visualization of the CMM concept in real-world intersections (upper part) and CMM-based co-simulation architecture (lower part).}
    \label{fig:Co-simulation Architecture}
\end{figure*}

As aforementioned, the cyber mobility mirror can further tap the potential of the traffic object surveillance systems to enable cooperative driving automation (CDA), especially for routing planning, cooperative decision-making, and motion control. From this perspective, this paper presents a co-simulation platform based on the CMM concept. Figure~\ref{fig:Co-simulation Architecture} demonstrates the concept of CMM with an intersection scenario and the system architecture of the CMM-based co-simulation.

In Figure~\ref{fig:Co-simulation Architecture}, the upper left part represents the real-world traffic scenario at an intersection, equipped with several roadside high-fidelity sensors, e.g., camera and 3D-LiDAR, roadside computing server, and communication system. High-fidelity sensor data is retrieved by the roadside server, in which perception tasks are executed, e.g., traffic object detection, classification, tracking, and motion prediction. Then perception results are encoded and transmitted to the CMM server via communication networks, e.g., cellular, DSRC or WLAN. The upper right part of Figure~\ref{fig:Co-simulation Architecture} represents the mirror environment, which reconstructs 3D objects based on perception information and outputs to CDA applications, e.g., collision avoidance, smart lane selection~\cite{jin2014improving}, and eco-driving~\cite{bai2022hybrid}, etc.

As aforementioned, building a comprehensive traffic object perception system in the real world requires plenty of hardware and labor resources. In this paper, we propose a cost-effective means to emulate the real-world traffic environment via a game-engine-based simulator, CARLA, which has the capability of generating traffic environment and high-fidelity sensor information. The structure of the real-world simulator, demonstrated in the lower left part of Figure~\ref{fig:Co-simulation Architecture}, consists of four modules: 1) initialization of system settings; 2) configuration of CARLA-based traffic environment with traffic objects and equipped sensors; 3) authentic perception using DNNs, such as ComplexYolo model; and 4) data encoding for communication. As shown in the lower right part of the figure, the Mirror Simulator is also developed based on CARLA and consists of three components: 1) decoding of the communicated data; 2) 3D rebuilding for vehicles and pedestrians, and 3) output interface for CDA applications to readily retrieve the post-perception data.
    
In this paper, we present the basic concept of CMM and design the concrete workflow of a roadside-sensor-based intersection surveillance scenario. In the following section of co-simulation design and development, we focus on: 1) applying roadside 3D-LiDAR for perception; 2) object detection and classification for vehicles and pedestrians; and 3) multi-vehicle 3D rebuilding for enabling CDA applications.

\subsection{Real-world Simulator Design}
The main purpose of the real-world simulator is to generate a virtual environment based on CARLA to emulate the real-world traffic environment. The main sub-tasks of the development effort are shown as follows.

\subsubsection{Traffic scenario design}
\begin{figure}[!h]
    \centering
    \includegraphics[width=0.5\textwidth]{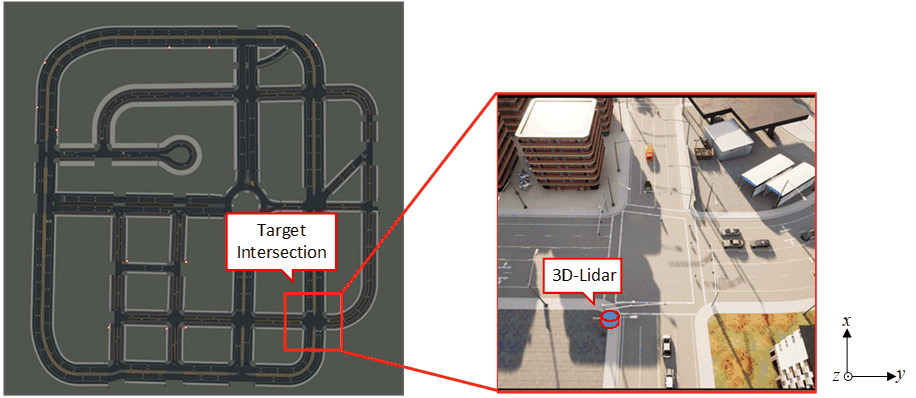}
    \caption{Traffic scenario design and 3D-LiDAR location.}
    \label{fig:carla town}
\end{figure}

CARLA has provided several well-developed virtual towns with different road maps and textures. In this paper, we implement ``town03" as our fundamental traffic map and select a target intersection for research. This is shown in the left part of Figure~\ref{fig:carla town}. Additionally, we can generate vehicles and pedestrians via CARLA-based python scripts. The right part of Figure~\ref{fig:carla town} shows a specific top-down view of the traffic scenario with vehicles, pedestrians, and a 3D-LiDAR installed at the target intersection. The traffic is generated via the CARLA interface with respect to certain traffic demands and the traffic signals are controlled by the built-in traffic signal manager in CARLA.

\subsubsection{Infrastructure based sensor design}
In this paper, we implement a roadside 3D-LiDAR as our main sensor. The roadside LiDAR is installed at the southwest corner of the target intersection, which is demonstrated in Figure~\ref{fig:carla town} (the LiDAR is installed below the arm of the traffic signal pole). Specifically, detailed settings about the roadside 3D-LiDAR are described later in Table~\ref{tab:parameter}. To reduce the concern on transferability of the deployed deep neural network model, we resemble the LiDAR setups as used to obtain the KITTI dataset \cite{6248074}. This means that the pitch, yaw, and roll of the LiDAR are set as zeros in CARLA global coordinate, as shown in the right part of Figure~\ref{fig:carla town}. Specifically, the LiDAR intensity is calculated by the following equation:
\begin{equation}
    I = I_{0} \cdot e^{-a\cdot d}
\end{equation}
where $I_{0}$ represents the initial intensity value (equals to 1 in this study); $a$ represents the attenuation coefficient, depending on the sensor's wavelength and atmospheric conditions (which can be modified by the LiDAR attribute “atmosphere\_attenuation\_rate”); and $d$ is the distance from the hit point to the sensor. Furthermore, realistic LiDAR features such as random no returns and background noise are also considered~\cite{thrun2005probabilistic}. More details about the implemented 3D-LiDAR are described in section~\ref{case study}.
    
\subsubsection{Deep Learning-Based Perception Methods}
\begin{figure}[!h]
    \centering
    \includegraphics[width=0.5\textwidth]{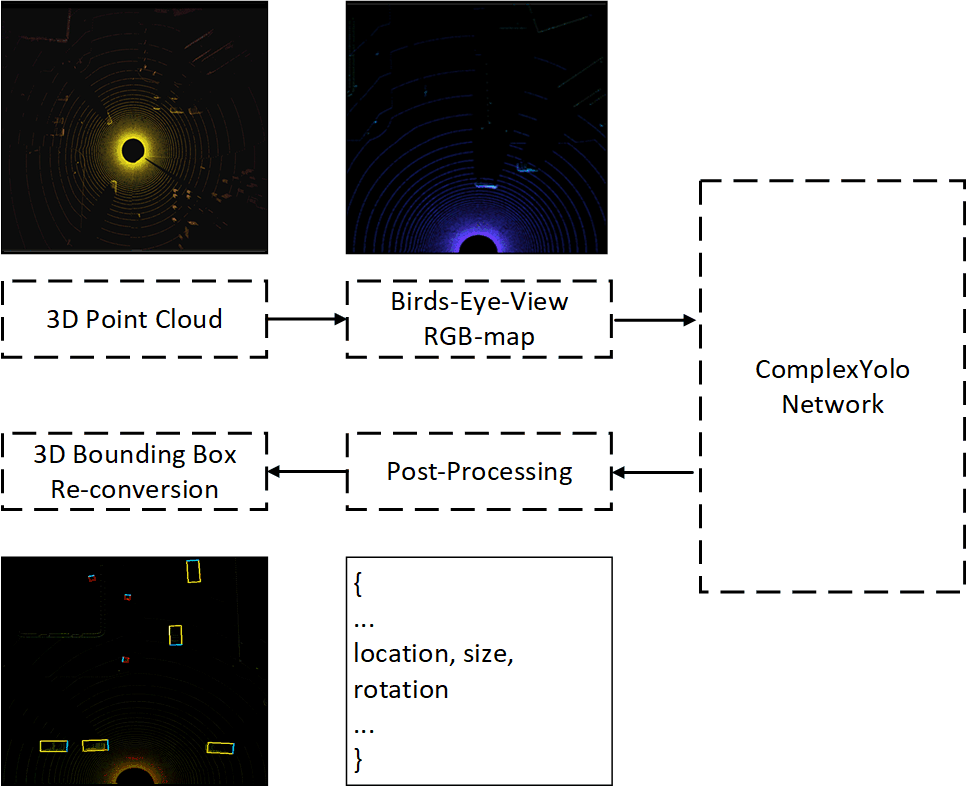}
    \caption{ComplexYolo-based real-time multi-class traffic object detection.}
    \label{fig:complexYolo}
\end{figure}

In this paper, we apply the ComplexYolo model \cite{simony2018complex} as our fundamental 3D object detection method. The basic pipeline of the ComplexYolo model is demonstrated in Figure~\ref{fig:complexYolo}.

In the ComplexYolo model, raw 3D-LiDAR data is firstly cut into a certain shape with respect to the target region; then the 3D point cloud is processed into Bird’s-Eye-View (BEV) RGB map based on different features; the CNN-based ComplexYolo network generates detection outputs based on the RGB map; moreover, post-processing is implemented to filter detection results with respect to certain thresholds; finally, 3D bounding boxes are calculated and displayed on the RGB map image. The optimization loss functions $L$ for ComplexYolo is defined as:
\begin{equation}
    L = L_{Yolo} + L_{Euler}
\end{equation}
where $L_{Yolo}$ is defined as the sum of squared errors using the multi-part loss introduced in YOLO \cite{Redmon_2016_CVPR} and YOLOv2 \cite{Redmon_2017_CVPR}, while the Euler regression part $L_{Euler}$ is defined to handle complex numbers, which has a closed mathematical space for angle comparisons \cite{simony2018complex}. The implementation details of the ComplexYolo model in this paper will be introduced in Section~\ref{case study}.

\subsection{Mirror Simulator Design}
The main purposes of the Mirror Simulator are: 1) performing 3D rebuilding for the perceived objects; and 2) providing a readily used output interface for CDA applications. 
    
In this paper, the Mirror Simulator is also developed based on the CARLA simulator. The same CARLA town, ``town03", is used to build this mobility mirror, as shown on the left in Figure~\ref{fig:carla town}. The 3D rebuilding procedure consists of two parts: 1) decoding post-perception data received from the real-world simulator via communication; and 2) generating 3D traffic objects with respect to the decoding data based on CARLA Python APIs (Application Programming Interfaces). Specifically, the message decoding method is further described in the next section.

\subsection{Co-Simulation Workflow Design}
In this paper, TCP protocol~\cite{2002TCP} is implemented to transmit post-perception data from the Real-world Simulator to the Mirror Simulator. Specifically, before transmitting, object detection results are encoded based on JSON protocol \cite{JoeLennon2009Introduction}, which includes the locations and orientations of the objects. For data encoding at the Mirror Simulator, the JSON data is decomposed according to the data structure in CARLA. 

\begin{figure*}[!ht]
    \centering
    \includegraphics[width=0.9\textwidth]{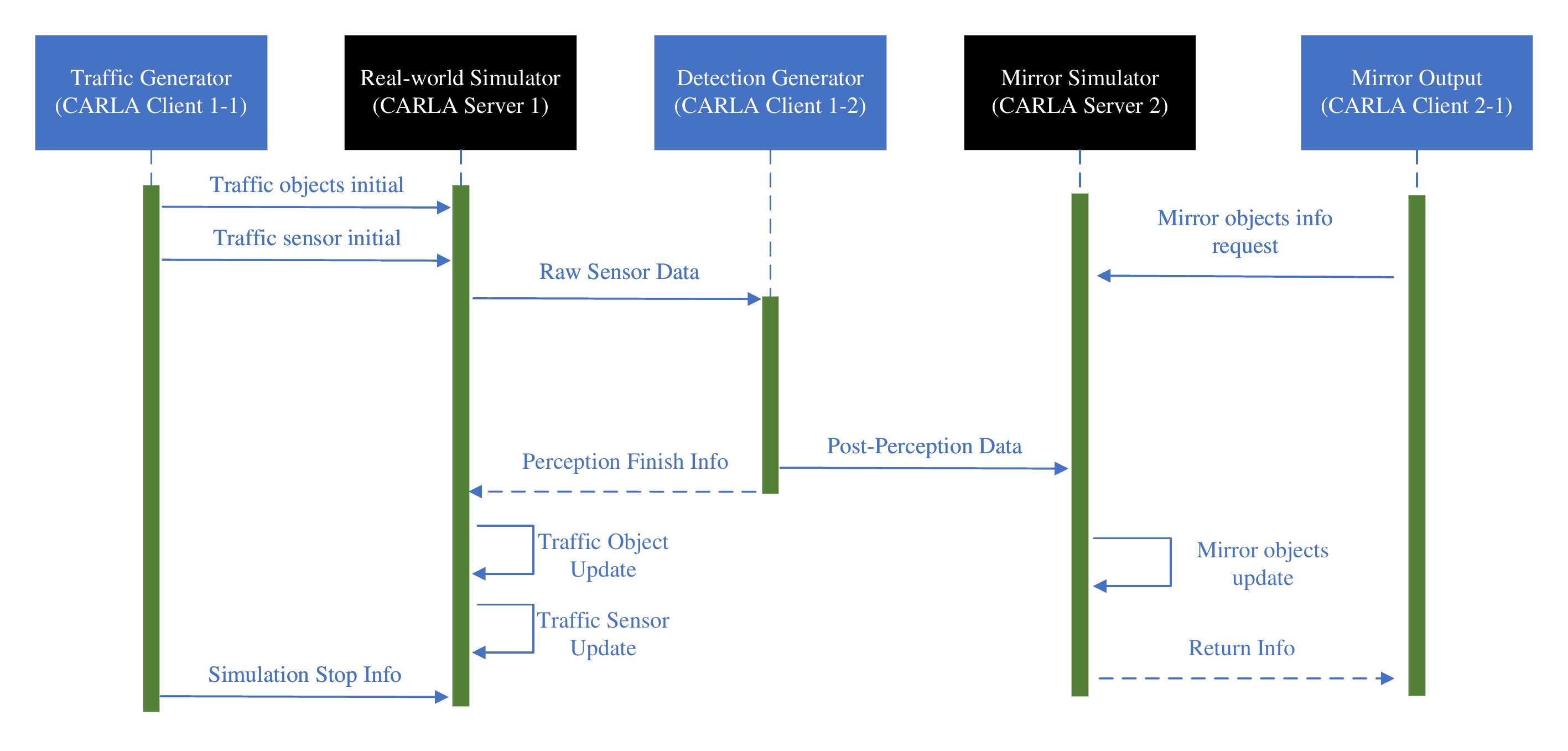}
    \caption{Sequence diagram for information synchronization among components.}
    \label{fig:sync}
\end{figure*}

For this co-simulation platform, we design the information flow based on the server-client architecture in CARLA \cite{dosovitskiy2017carla}. The sequence diagram for information synchronization among components is demonstrated in Figure~\ref{fig:sync}. 

In the CARLA platform, the server runs the simulation (i.e., updates the information), while the client retrieves information. Specifically, as shown in Figure~\ref{fig:sync}, the Traffic Generator (CARLA Client 1-1) is responsible for simulation initialization or stopping requests; the Real-world Simulator (CARLA Server 1) runs the simulation for real-world traffic and the LiDAR sensor; the Detection Generator (CARLA Client 1-2) is designed to generate 3D object detection results by utilizing the ComplexYolo model. The detection results are encoded into post-perception data and transmitted to the Mirror Simulator (CARLA Server 2) via TCP communications. Finally, the Mirror Output (CARLA Client 2-1) can retrieve the mirrored objects’ information, e.g., vehicle center location, bounding box dimension, and orientation, from the Mirror Simulator.
\subsection{V2X Effect Design}
\label{v2x}
Since Vehicle-to-everything (V2X) communication effects are significant in CDA implementations, this section introduces the V2X effect design from three different perspectives.
\subsubsection{Communication Delay}
Owing to the co-simulation structure, the communication delay is naturally involved in deploying Realworld Simulation and Mirror Simulation on different computers. Furthermore, to support assessment for different delay conditions, an active communication delay (ACD) is designed at Detection Generator shown in Figure~\ref{fig:sync}. ACD aims to provide customized time delays that can be defined according to specific requirements. Hence, the total delay $\mathcal{T}$ of the co-simulation is designed as:
\begin{equation}
    \mathcal{T} = \mathcal{T}_{In} + \mathcal{T}_{ACD}
\end{equation}
where $\mathcal{T}_{In}$, and $\mathcal{T}_{ACD}$ represent the innate time delay of the co-simulation and the delay from ACD. Specifically, for wireless communication applied in the our case study shown in Section~\ref{case study}, $\mathcal{T}_{In}$ is about $150ms$ and $\mathcal{T}_{ACD} \sim \mathcal{N}(\mu, \sigma^{2})$ is applied by setting $\mu, \sigma$ as $50$ and $5$ respectively.

\subsubsection{Message Dropping}
Considering that message dropping is a common issue in communications systems~\cite{krifa2011message}, an Active Message Dropping (AMD) mechanism is designed in our co-simulation system. In the Co-simulation system, AMD is designed by the stochastic process and deployed right after the ACD module. The $\mathcal{R}$ represents the factor for the message dropping, which is designed by:
\begin{equation}
    \mathcal{R} \sim \mathcal{U}(0, 1)
\end{equation}
For each frame, if $\mathcal{R}$ is smaller than a certain threshold $\eta$, the message of this frame will be drooped. Specifically, for experiments in section~\ref{functionCDA}, $\eta$ is set as $0.0$, $0.05$, and $0.10$ respectively.

\section{Case Study: Roadside-LiDAR-Based Vehicle Detection For Enabling CDA}
\label{case study}
\subsection{System Setup}
Both the Real-world Simulator and Mirror Simulator run under the synchronous mode of CARLA, which means the server can update the simulator information at the same time step with the clients. In this paper, the simulation frequency is set as 10 Hz. The network traffic demand is set to be 100 vehicles (driving around the town according to the default routing strategy and autopilot method in CARLA) \cite{dosovitskiy2017carla}. For the 3D-LiDAR implemented in this paper, attributes of the LiDAR sensor are shown in Table~\ref{tab:parameter}.

\begin{table*}[!ht]
\centering
\caption{Parameter Configuration and Description for the 3D-LiDAR sensor deployed in the case study.}
\label{tab:parameter}
\resizebox{0.95\textwidth}{!}{%
\begin{tabular}{l|l|lll}
\cline{1-3}
Parameters          & Default & Description                                      &  &  \\ \cline{1-3}
Channels            & 64      & Number of lasers.                                 &  &  \\
Height              & 1.73$m$   & Height with respect to the road surface.        &  &  \\
Range               & 100.0$m$  & Maximum distance to   measure/ray-cast in meters. &  &  \\
Rotation\_frequency & 10.0$Hz$ & LiDAR rotation frequency.                         &  &  \\
Upper\_fov          & 2.0     & Angle in degrees of the highest   laser beam.         &  &  \\
Lower\_fov          & -24.9   & Angle in degrees of the lowest   laser beam.          &  &  \\
Atmosphere\_attenuation\_rate & 0.004   & Coefficient that measures the LiDAR intensity loss.                                           &  &  \\
Noise\_stddev                 & 0.01 & The standard deviation of the noise model of points along its ray-cast. &  &  \\ 
Dropoff rate                 & 45\% & General proportion of points that are randomly dropped. &  &  \\ 
Dropoff\_intensity\_limit      & 0.8 & The threshold of intensity value for exempting dropoff. &  &  \\ 
Dropoff\_zero\_intensity      & 40\% & The probability value of dropoff for zero-intensity points. &  &  \\ 
\cline{1-3}
\end{tabular}%
}
\end{table*}

\subsection{CARTI Dataset}
To implement the ComplexYolo model in our co-simulation platform, a well-defined 3D-LiDAR dataset with the ground-truth label is required. Although CARLA provides a comprehensive Python API for data retrieving and object controlling, a built-in dataset generator is still missing. Therefore, based on the existing CARLA Python API and the KITTI dataset structure, we develop a CARLA-based 3D-LiDAR Dataset which is named \textit{CARTI} Dataset. The code for generating the CARTI dataset is available at \textit{\url{https://github.com/zwbai/CARTI_Dataset}}. Although the LiDAR is running in $10Hz$, the data frame is recorded at $2Hz$ to improve the diversity of the dataset, i.e., including more differences in certain frames of data. Specifically, a total of 11,000 frames of data are collected.

\subsection{Vehicle Detection}
\subsubsection{Point Cloud Preprocessing}
In this paper, our target range for vehicle detection is defined as a 50 meters by 50 meters area $\Omega$ with respect to the location of LiDAR. The square size of the target range is due to the construction mechanism of the BEV feature map which is shown in section~\ref{bev}. The raw point cloud data can be described by:
\begin{equation}
    \mathcal{P} = \{[x, y, z, i]\, |\, [x, y, z] \in \mathbb{R}^{3}, i\in [0.0, 1.0] \}.
\end{equation}
To reduce the impact from 3D LiDAR points out of the target range, $\mathcal{P}$ is geo-fenced by:
\begin{equation}
\mathcal{P}_{\Omega} = \{[x, y, z, i]^{T} \, |\, x \in \mathcal{X}, y \in \mathcal{Y}, z \in \mathcal{Z}\}   
\end{equation}
where $\mathcal{P}_{\Omega}$ represents the 3D point cloud data after geofencing; and $\mathcal{X}$ and $\mathcal{Y}$ are set as $[0.0m, 50.0m], [-25.0, 25.0]$ respectively. Considering the calibrated height of the roadside Lidar to be $1.74m$, $\mathcal{Z}$ is set as $[-2.74m, 1.36m]$. Additionally, ground truth labels are collected according to the objects within the target range. 

\subsubsection{BEV Image Construction}
\label{bev}
The 3D points within the target range are then normalized by the following equations:
\begin{equation}
\begin{bmatrix}
\tilde{x}\\ 
\tilde{y}\\ 
\tilde{z}
\end{bmatrix} =\begin{bmatrix}
\frac{h}{range_x} &  0& 0\\ 
0 &  \frac{h}{range_y}&0 \\ 
 0&  0& 1
\end{bmatrix}\begin{bmatrix}
x\\ 
y\\ 
z
\end{bmatrix}+\begin{bmatrix}
0\\ 
0.5w\\ 
0
\end{bmatrix}
\end{equation}
where $range_x$ and $range_y$ represent the range along the x-axis direction and y-axis direction, respectively; $w$ and $h$ denote the weight and height of the BEV image, respectively. Specifically, $[x\ y\ z]^T$ and $[\tilde{x}\ \tilde{y}\ \tilde{z}]^T$ represent the points in $\mathcal{P}_{\Omega}$ and the points after normalization respectively.

Then the three-feature maps, i.e., R-map, G-map, and B-map can be defined according to the density, height, and intensity information, respectively, which is shown below:
    
\begin{equation}
\begin{matrix}
z_{r}(S_{j}) = min(1.0, \log(N+1)/64, N=|P_{\Omega\,\ i\rightarrow j}|)\\
z_{g}(S_{j}) = max(P_{\Omega\,\ i\rightarrow j} \cdot [0, 0, 1]^T)\\ 
z_{b}(S_{j}) = max(I(P_{\Omega\,\ i\rightarrow j}))
\end{matrix}
\end{equation}
where $S_j$ represents a specific grid cell of RGB-map; $z_r$ , $z_g$  and $z_b$ represent three channels for RGB-map; $I$ represents the intensity of LiDAR point and $N$ describes the number of points mapped from $P_{\Omega\,\ i}$ to $S_j$.

\subsubsection{Model Training}
Based on the CARTI dataset, we train the ComplexYolo model from scratch via stochastic gradient descent with a weight decay of 0.0005 and momentum of 0.9. For the dataset preparation, we subdivide the training set with 80\% for training and 20\% for validation. The learning rate is set as 0.001 for initialization and gradually decreased along 1000 training epochs. For regularization, we implement batch normalization. For activation functions, a leaky rectified linear activation function, defined as follows, is used except in the last layer of the convolution neural network (CNN) where a linear activation function $f(x) = x$ is used.

\begin{equation}
f(x) = \left\{\begin{matrix}
x, & x> 0\\ 
0.1x, & otherwise
\end{matrix}\right.
\end{equation}

\subsection{Functionality for Authentic Perception}
\subsubsection{Quantitative Performance}

\begin{figure}[!ht]
    \centering
    \includegraphics[width=0.5\textwidth]{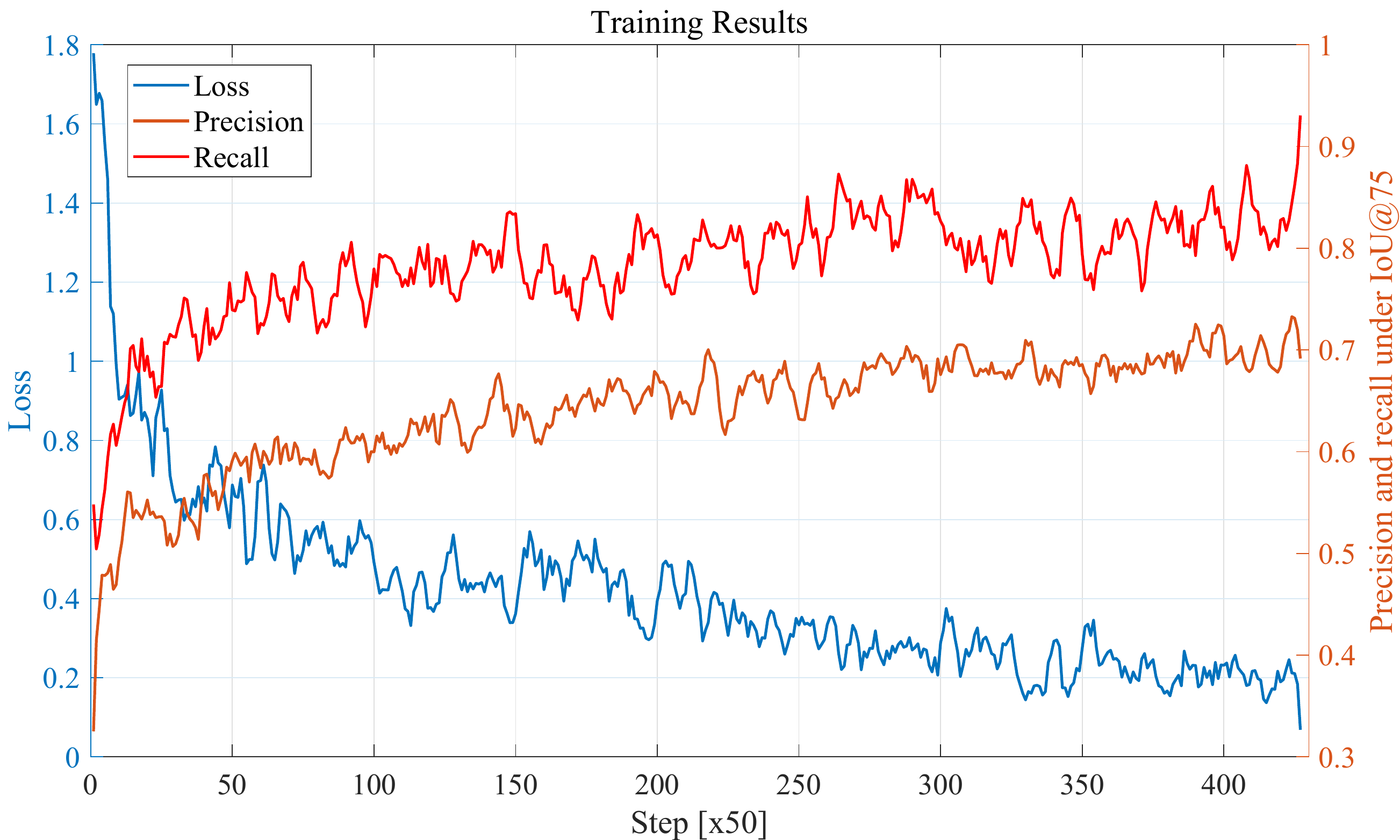}
    \caption{Training performance based on CARTI Dataset.}
    \label{fig:train-1}
\end{figure}

For quantitative results and analysis, \cref{fig:train-1} demonstrate the overall performance of the training results including the training loss, precision, and recall. Specifically, \textit{Precision} illustrates the proportion of true positive detection among all predicted detection (i.e., all the bounding boxes generated from the model). \textit{The recall} is a factor that measures the ability of the detector to find all the relevant cases (i.e., all the ground truths) which is the proportion of true positive detection among all ground truths (i.e., real vehicles and pedestrians). Additionally, the notations of $@50$ and $@75$ mean these values are calculated based on the Intersection of Union (IoU) of $0.50$ and $0.75$,  respectively. 


The evaluation results of the testing dataset are demonstrated in Table~\ref{tab:eva}. Specifically, \textit{AP} represents the average precision for each class along the different thresholds and \textit{F1} is the harmonic mean for precision and recall, which is designed as:

\begin{equation}
    F1 = 2\cdot\frac{Precision \cdot Recall}{Precision +  Recall}
\end{equation}

It is notable that the detection results for vehicles are much better than for pedestrians. A hypothesis is that fewer LiDAR points would be reflected from pedestrians than vehicles and pedestrians are more susceptible to occlusion, due to their smaller sizes.

\begin{table}[!h]
\centering
\caption{Details about the evaluation results under IoU$@75$ (in \%)}
\label{tab:eva}
\resizebox{0.45\textwidth}{!}{%
\begin{tabular}{llllll}
\hline
\multicolumn{2}{l}{Evaluation Criteria}                                              & Precision & Recall & AP    & F1    \\ \hline
\multicolumn{1}{l|}{\multirow{2}{*}{Class}} & \multicolumn{1}{l|}{Car}        & 84.85     & 95.93  & 95.41 & 90.05 \\ \cline{2-6} 
\multicolumn{1}{l|}{}                              & \multicolumn{1}{l|}{Pedestrian} & 35.71     & 42.00  & 21.45 & 38.60 \\ \hline
\end{tabular}%
}
\end{table}

\begin{figure}[!b]
    \centering
    \includegraphics[width=0.48\textwidth]{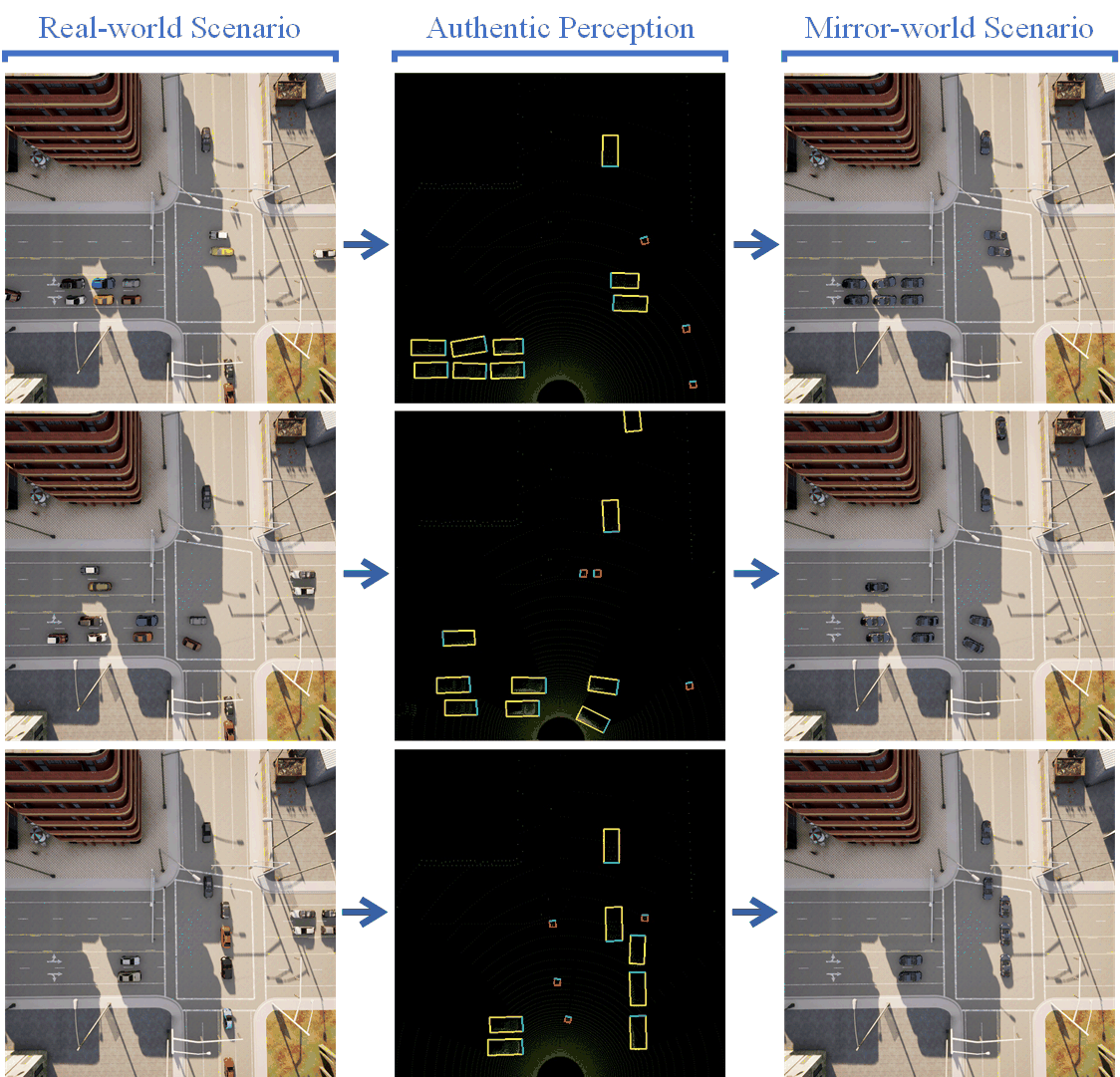}
    \caption{Visualization of object detection results: 1) top-down view of the ``real-world'' traffic environment (left-column images), 2) visualization for the detection results (mid-column images), and 3) the top-down view of the ``mirror world'' traffic conditions (right-column images).}
    \label{fig:evaluation}
\end{figure}

\subsubsection{Qualitative Performance}
For qualitative results and analysis, Figure~\ref{fig:evaluation} demonstrates the key pipeline for the co-simulation system. The left-column images show the top-down view of the ``real-world'' traffic environment, the authentic perception results are visualized in mid-column images, and the traffic conditions in the ``mirror world'' are demonstrated by the right-column top-down view images. For detection visualization, vehicles and pedestrians are bounded with yellow and red boxes, respectively. In addition, the blue edges of the bounding boxes represent the forward direction of the objects.

Figure~\ref{fig:evaluation} validates the feasibility of our CMM-based co-simulation platform. Vehicles and pedestrians are detected via the roadside-based LiDAR and the associated digital replica is reconstructed in the mobility mirror simulator. Due to the detection range and accuracy of the selected model (i.e., ComplexYolo), some objects will be missed. Nevertheless, our platform is generic and highly compatible with different detection models and the results can be improved with the advances in SOTA detection methods.

In general, Figure~\ref{fig:evaluation} validates the core concept of our CMM-based co-simulation platform, i.e., generating authentic detection results based on high-fidelity sensors and rebuilding the traffic objects for external CDA applications, which demonstrates the functionality and feasibility of the co-simulation platform.

\subsection{Functionality for Enabling CDA}
\label{functionCDA}
\begin{figure}[!h]
    \centering
    \includegraphics[width=0.48\textwidth]{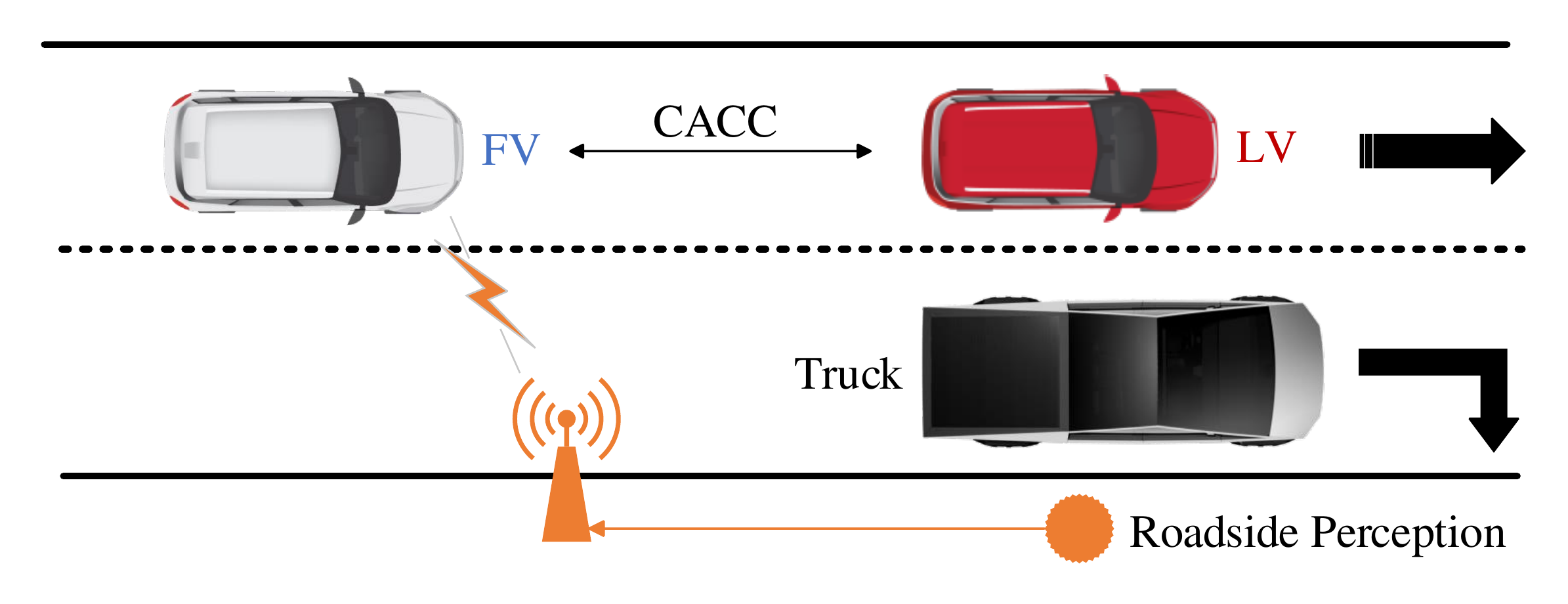}
    \caption{A case scenario of CDA in mixed traffic enabled by our platform.}
    \label{fig:eva-cacc}
\end{figure}

This section will demonstrate two main aspects of the functionalities of our co-simulation platform in terms of enabling CDA. As shown in Figure~\ref{fig:eva-cacc}, a specific CDA application -- infrastructure-assisted CACC -- is designed in a mixed traffic scenario to illustrate 1) why authentic perception is important for supporting CDA algorithm development and evaluation; and 2) how can the co-simulation platform supports the CDA applications.

In Figure~\ref{fig:eva-cacc}, the ego-vehicle (marked as FV, i.e., the following vehicle), a CV without self-perception capacity, is trying to follow a leading vehicle (LV), a legacy vehicle without connectivity. The location and speed information of the leading vehicle is required for this scenario. 

\begin{figure}[!t]
    \centering
    \includegraphics[width=0.48\textwidth]{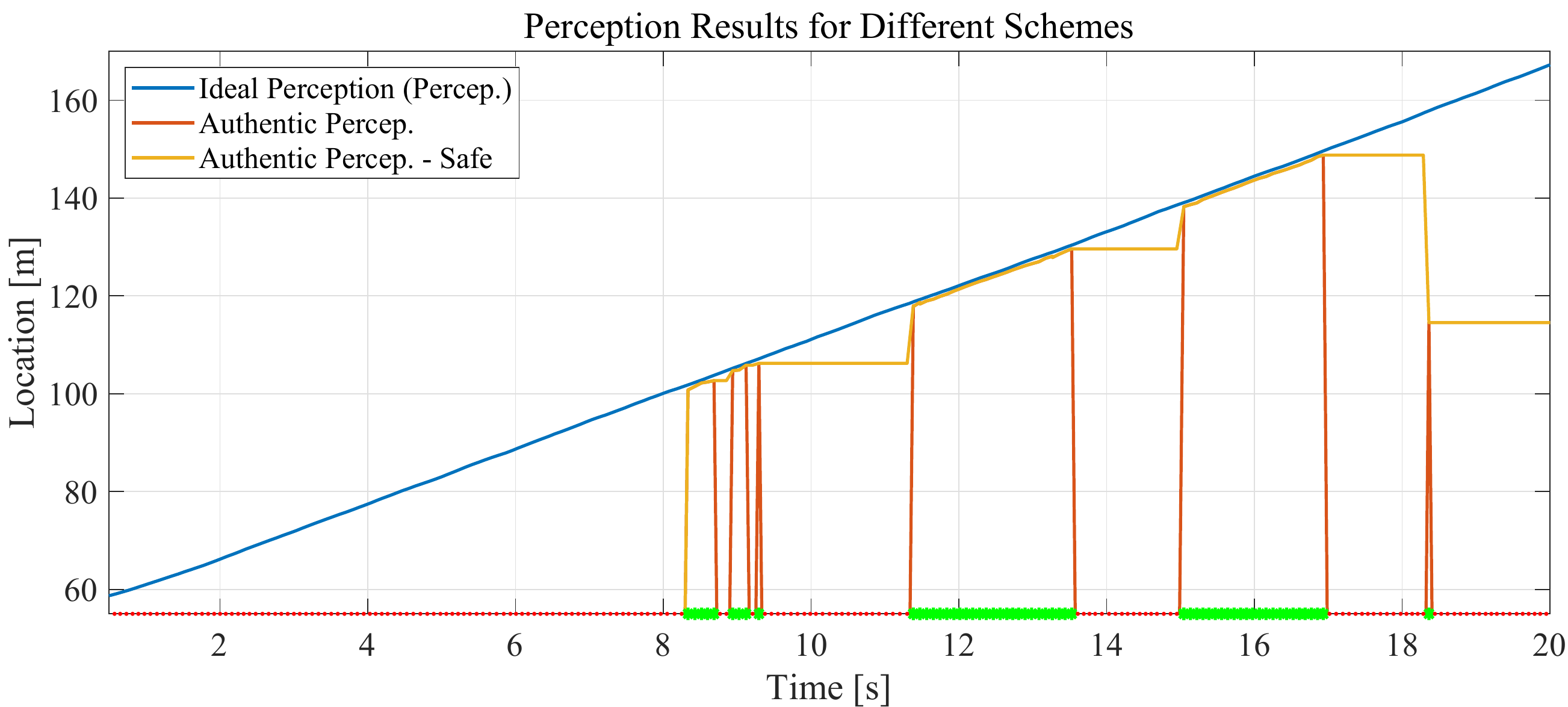}
    \caption{Perception results from different perception schemes.}
    \label{fig:eva-perception}
\end{figure}

\begin{figure}[!t]
    \centering
    \includegraphics[width=0.48\textwidth]{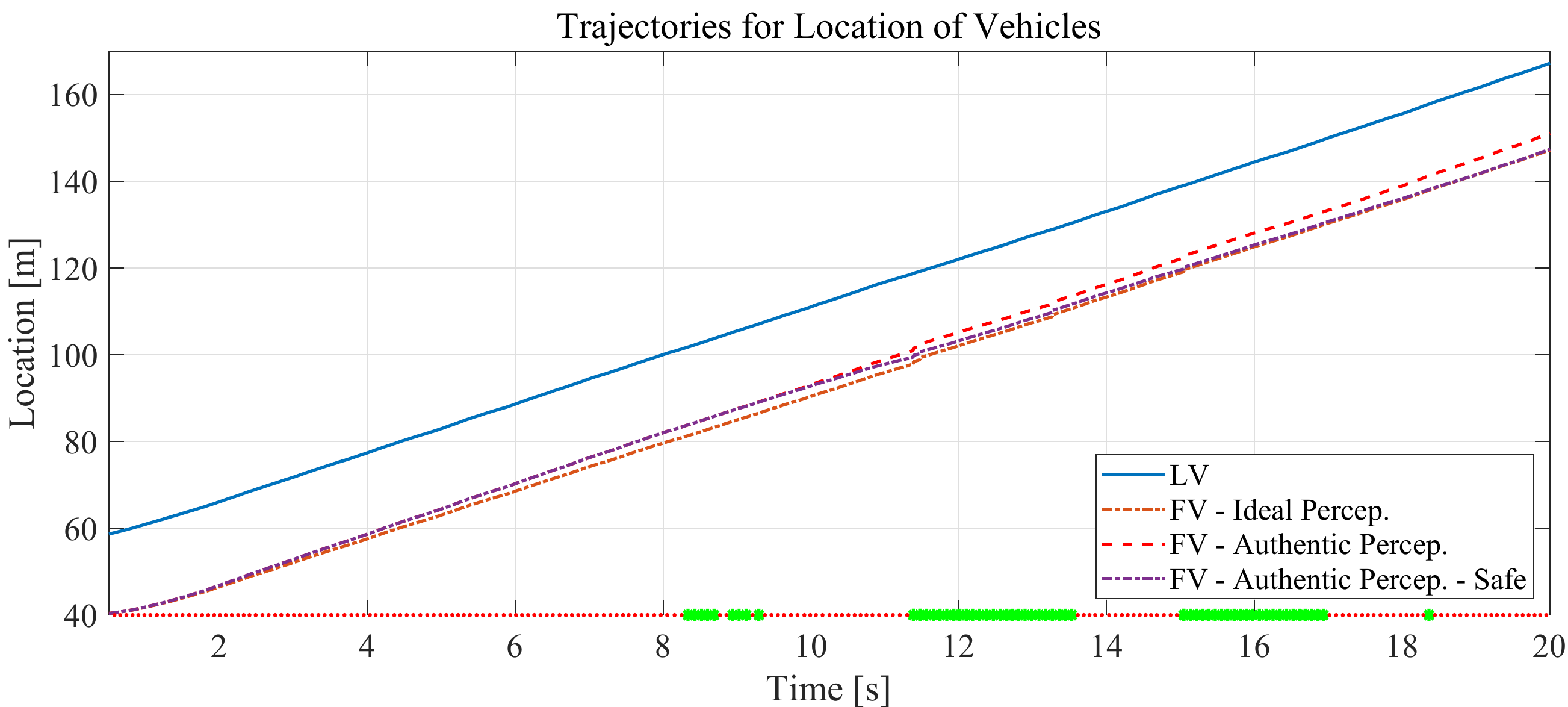}
    \caption{Trajectories for location based on different perception results.}
    \label{fig:eva-location}
\end{figure}

\begin{figure}[!t]
    \centering
    \includegraphics[width=0.48\textwidth]{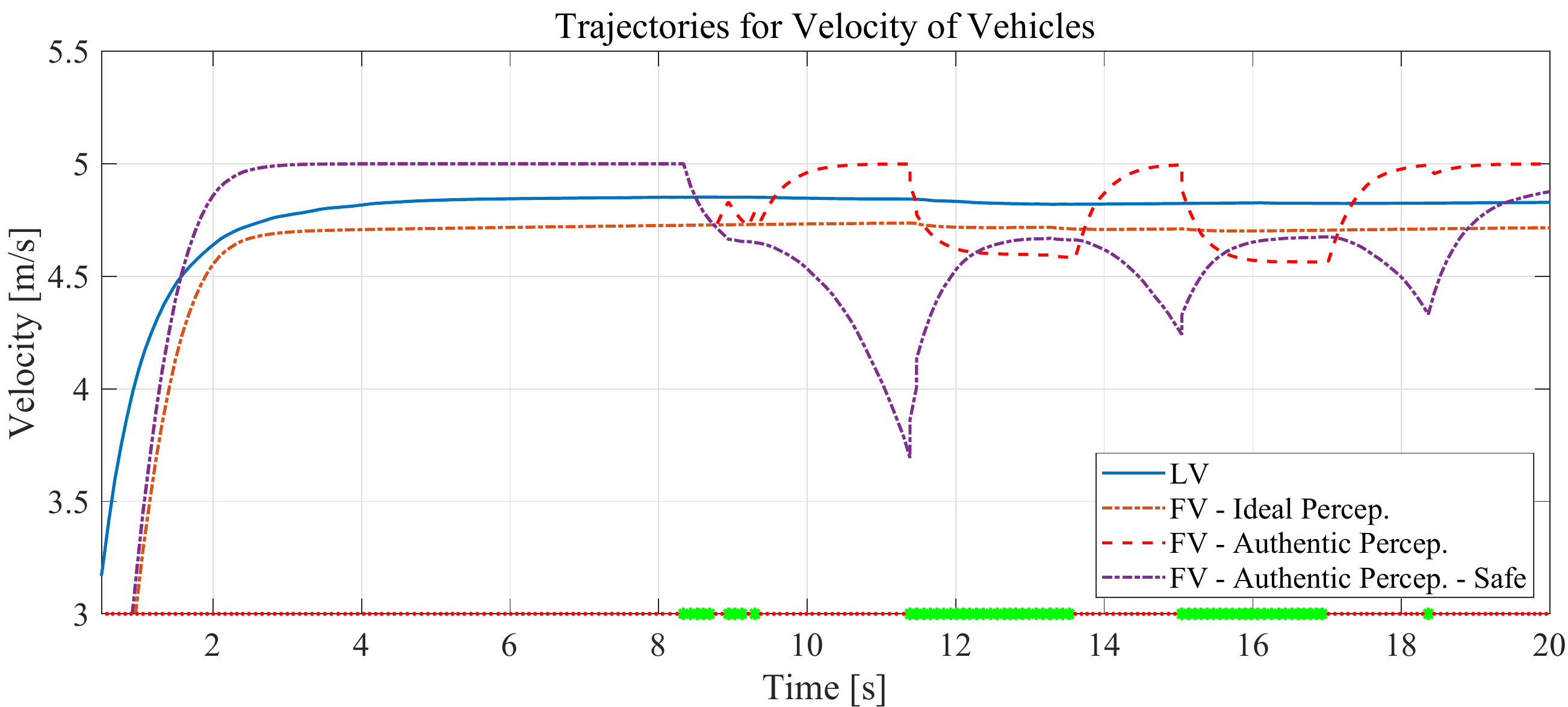}
    \caption{Trajectories for velocity based on different perception results.}
    \label{fig:eva-velocity}
\end{figure}

\begin{figure}[!t]
    \centering
    \includegraphics[width=0.48\textwidth]{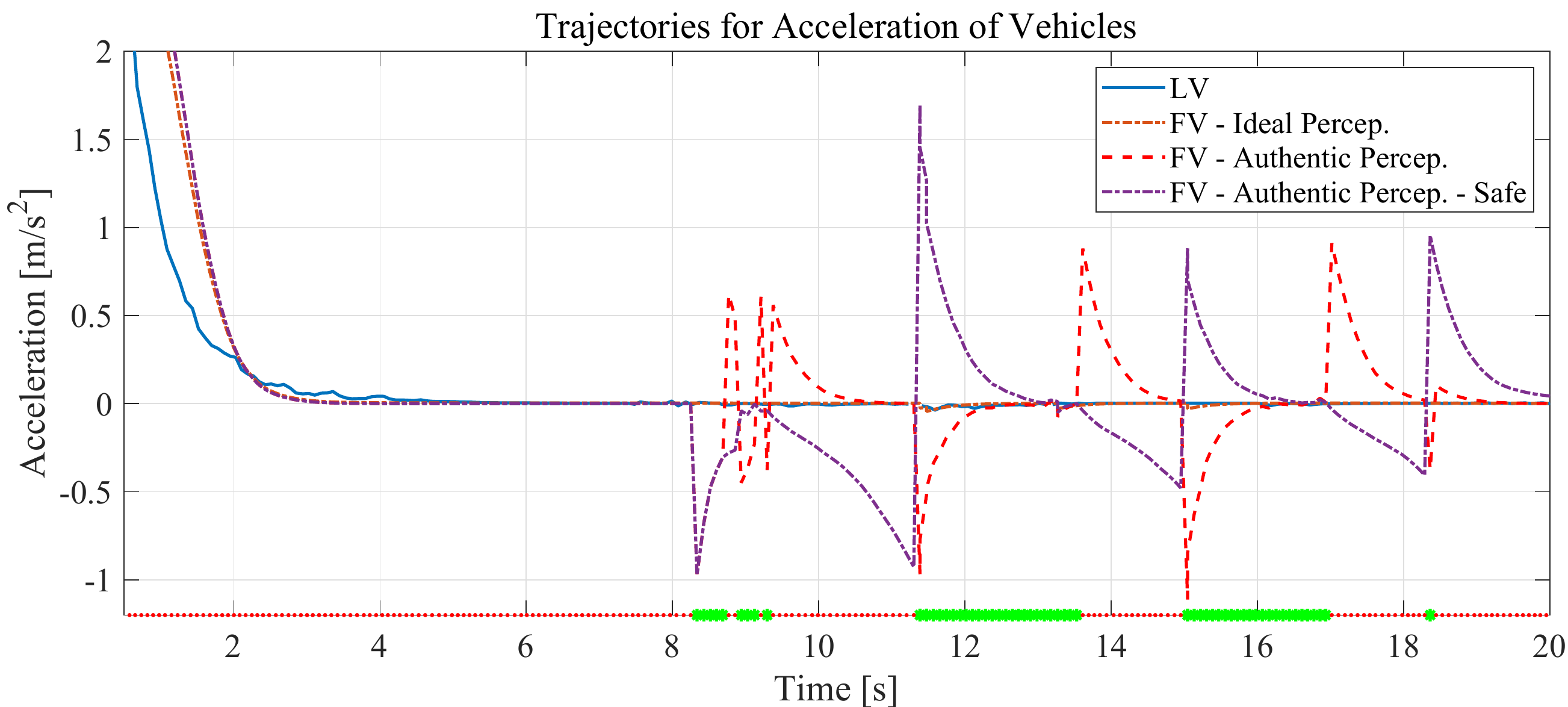}
    \caption{Trajectories for acceleration based on different perception results.}
    \label{fig:eva-acc}
\end{figure}

In this paper, different perspectives are considered to make the assessment more comprehensive and realistic: 1)mobile occlusion, e.g., LV in Figure~\ref{fig:eva-cacc} is occluded by the Truck, 2) time delay, and 3) message drop. To be specific, the Intelligent Driver Model (IDM) is applied to act as a basic CACC method, and when miss detection happens, two schemes are designed  to generate continuous perception results: 1) there are no leading vehicles; and 2) the LV will keep its location as before. Since scheme \#1 represents a progressive CACC style, scheme \#2 represents a conservative CACC style. As demonstrated by Figure~\ref{fig:eva-perception}, Ideal Perception (IP) represents the ground truth vehicle location, and Authentic Perception (AP) and AP-Safe (AP-S) are the visualizations for the two schemes mentioned above. Particularly, the time when detection of LV is received is marked by a green ``\textcolor{green}{$*$}'' at the bottom of the figures while a red  ``\textcolor{red}{$\cdot$}'' is marked for miss detection happens. Based on the three perception results, vehicle trajectories are generated and shown in Figures ~\ref{fig:eva-location} to~\ref{fig:eva-acc}. In Figure~\ref{fig:eva-location}, FVs based on different perception results can keep a reasonable distance gap to the LV. In detail, the trajectory of AP-S has a closer fit with the trajectory of Ideal Perception. 


Although three perception schemes can support FV generating feasible location trajectories, the speed and acceleration trajectories generated from authentic perception (i.e., AP and AP-S) have significant differences from the trajectory generated by ideal perception (i.e., IP), which are shown in Figure~\ref{fig:eva-velocity} and \ref{fig:eva-acc}. For velocity, trajectories generated by AP and AP-S have much more speed fluctuations which are mainly caused by 1) the miss detection caused by occlusion and 2) wrong detection caused by the performance of the detection model itself, such as the detection at $18.4s$ shown in Figure~\ref{fig:eva-perception}. Thus, it is quite difficult to model the deficiency of authentic perception in a traditional way, such as the probabilistic model. 
For acceleration, a similar pattern can be found in Figure~\ref{fig:eva-acc}. The trajectories based on AP and AP-S have more fluctuations compared with the trajectory of IP. 

Furthermore, it is notable that the fluctuations are highly correlated with the status of perception results. Specifically, AP will accelerate will the LV is missing and AP-S tends to decelerate until the LV is detected again. These trajectory patterns illustrate that the perception process and results will significantly affect the performance of CACC in this case or similar to other CDA applications, which further demonstrates the necessity to take authentic perception into consideration when designing subsequent CDA algorithms.

\begin{figure}[!t]
    \centering
    \includegraphics[width=0.48\textwidth]{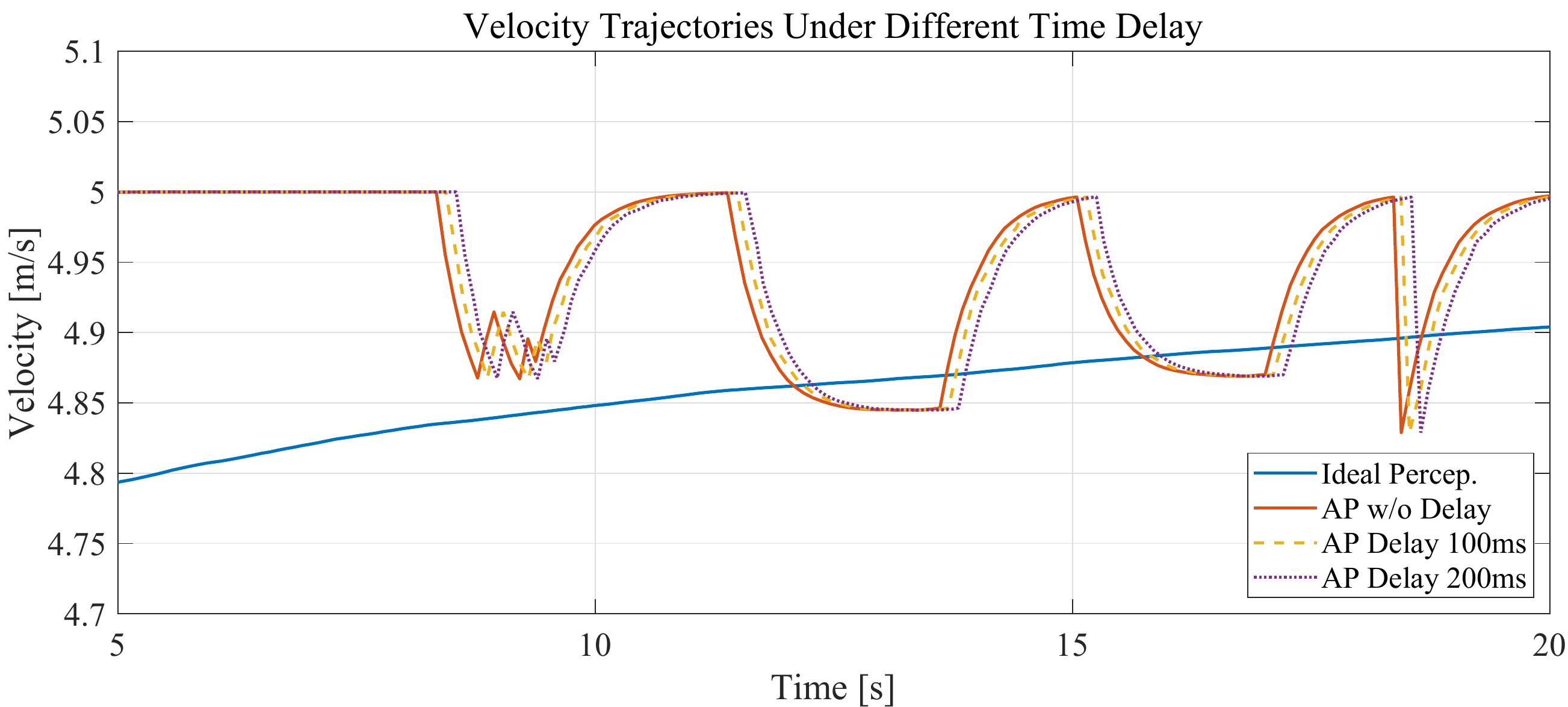}
    \caption{Velocity trajectories based on different perception delays.}
    \label{fig:eva-delay}
\end{figure}
\begin{figure}[!t]
    \centering
    \includegraphics[width=0.48\textwidth]{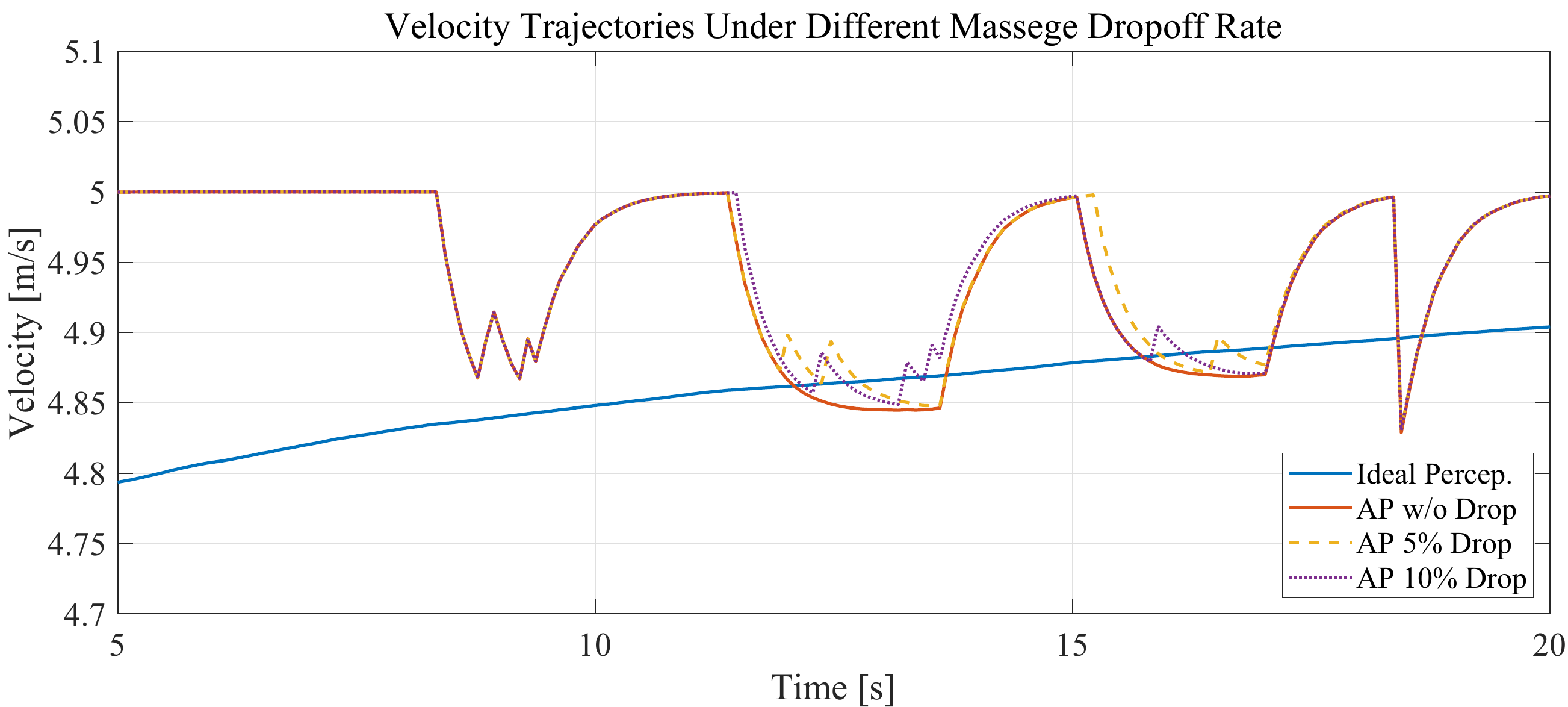}
    \caption{Velocity trajectories based on different perception message drop.}
    \label{fig:eva-drop}
\end{figure}

For the functionality of our platform to support the V2X effect, time delay and message drop are considered in section~,\ref{v2x}, and experiments are conducted accordingly in this section. Figure~\ref{fig:eva-delay} shows the velocity trajectories based on IP, AP without (w/o) delay, AP with $100ms$ delay, and AP with $200ms$ delay, respectively. Figure~\ref{fig:eva-drop} shows the velocity trajectories based on IP, AP without message drop (w/o drop), AP with 5\% message drop, and AP with 10\% message drop, respectively. Results from Figures ~\ref{fig:eva-delay} and~\ref{fig:eva-drop} demonstrate that our platform can support the analysis of the V2X effect which plays a crucial role in CDA implementation.

\section{Conclusion and Discussion}
\label{conclusion}
To enable cooperative driving automation (CDA), simulators are imperative to comprehensively support the design and assessment of various applications. Moreover, authentic perception based on high-resolution sensors is of great significance for CDA development. To the best of the authors’ knowledge, this paper is the first attempt to design and develop a co-simulation platform to prove the cyber mobility mirror (CMM) concept, which can both emulate high-resolution sensors and provide readily retrieved perception information. Specifically, the co-simulation platform consists of two main sub-simulators: 1) the Real-world Simulator for emulating the real-world traffic environment and (roadside) sensors and generating the authentic perception data; and 2) the Mirror Simulator for 3D reconstructing traffic objects and providing a readily retrieved interface for downstream CDA applications to access the location and orientation information of target traffic objects. A case study is conducted for roadside LiDAR-based vehicle detection in an intersection scenario, which demonstrates the performance of authentic perception as well as the functionality and feasibility of the co-simulation platform for enabling CDA.
    
In this paper, we develop a preliminary framework for CMM and validate it with simulation. A natural future step would be to realize the system in the real world, but there are some open issues deserving further exploration. For instance, we need to overcome disparities in the features between sensor data from simulators and that in reality. We need to investigate the model transferability issue, i.e., to design a model that can be trained on simulation and implemented in real-world scenarios without the necessity or much effort in re-training the model or fine-tuning parameters.

\section*{Acknowledgments}
This research was funded by the Toyota Motor North America InfoTech Labs. The contents of this paper reflect the views of the authors, who are responsible for the facts and the accuracy of the data presented herein. The contents do not necessarily reflect the official views of Toyota Motor North America.

\bibliographystyle{IEEEtran}
\typeout{}
\bibliography{References}{}


\begin{IEEEbiography}
    [{\includegraphics[width=1in,height=1.25in,clip,keepaspectratio]{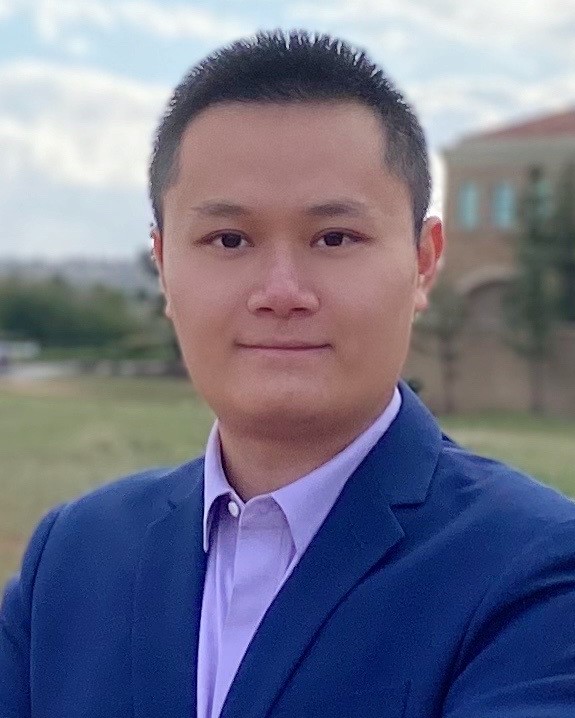}}]{Zhengwei Bai}
(Student Member, IEEE) received the B.E. and M.S. degrees from Beijing Jiaotong University, Beijing, China, in 2017 and 2020, respectively. He is currently a Ph.D. candidate in electrical and computer engineering at the University of California at Riverside. His research focuses on object detection and tracking,  cooperative perception, decision making, motion planning, and cooperative driving automation (CDA). He serves as a Review Editor in Urban Transportation Systems and Mobility.
\end{IEEEbiography}

\begin{IEEEbiography}
    [{\includegraphics[width=1in,height=1.25in,clip,keepaspectratio]{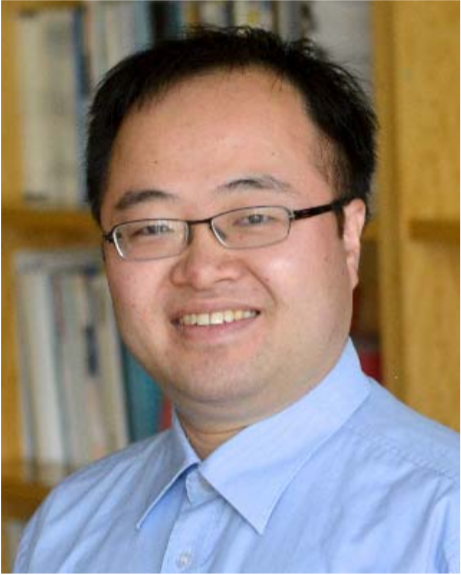}}]
    {Guoyuan Wu}
(Senior Member, IEEE) received his Ph.D. degree in mechanical engineering from the University of California, Berkeley in 2010. Currently, he holds an Associate Researcher and an Associate Adjunct Professor position at Bourns College of Engineering – Center for Environmental Research \& Technology (CE–CERT) and the Department of Electrical \& Computer Engineering at the University of California at Riverside. development and evaluation of sustainable and intelligent transportation system (SITS) technologies, including connected and automated transportation systems (CATS), shared mobility, transportation electrification, optimization and control of vehicles, traffic simulation, and emissions measurement and modeling. Dr. Wu serves as Associate Editors for a few journals, including IEEE Transactions on Intelligent Transportation Systems, SAE International Journal of Connected and Automated Vehicles, and IEEE Open Journal of ITS. He is also a member of the Vehicle-Highway Automation Standing Committee (ACP30) of the Transportation Research Board (TRB), a board member of the Chinese Institute of Engineers Southern California Chapter (CIE-SOCAL), and a member of the Chinese Overseas Transportation Association (COTA). He is a recipient of the Vincent Bendix Automotive Electronics Engineering Award.
\end{IEEEbiography}

\begin{IEEEbiography}
    [{\includegraphics[width=1in,height=1.25in,clip,keepaspectratio]{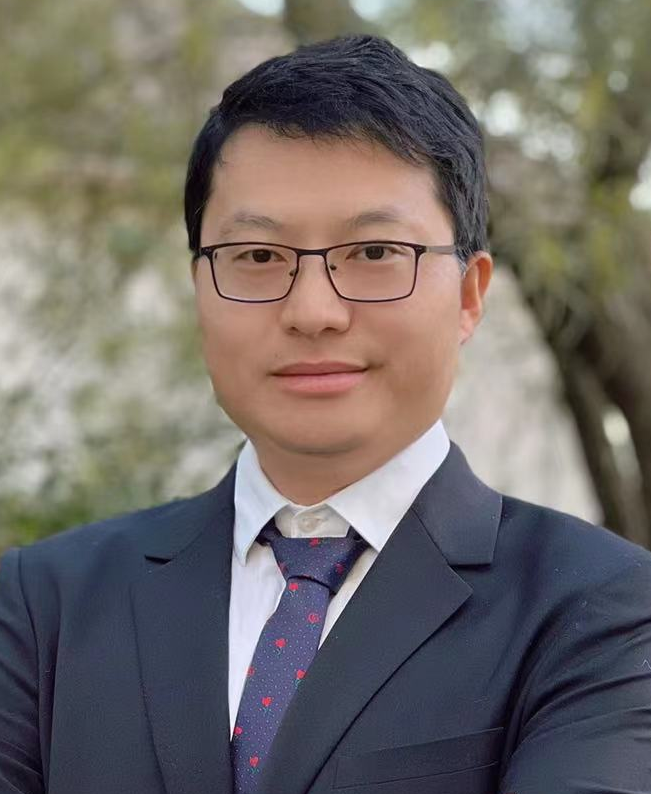}}]
    {Xuewei Qi}
(Member, IEEE) received his Ph.D. degree in electrical and computer engineering from the University of California-Riverside in 2016 and his M.S. degree in engineering from the University of Georgia, USA, in 2013. He is a Principle AI Researcher with Toyota North America Research Labs (Silicon Valley). He was with General Motors as an Artificial Intelligence and Machine Learning Research Scientist. He was also working as a Lead Perception Research Engineer at Aeye.ai. His recent research interests include deep learning, autonomous vehicles, perception and sensor fusion, reinforcement learning, and decision making. He is also serving as a member of several standing committees of the Transportation Research Board (TRB).
\end{IEEEbiography}
\begin{IEEEbiography}
    [{\includegraphics[width=1in,height=1.25in,clip,keepaspectratio]{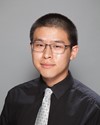}}]{Yongkang Liu}
received the Ph.D. and M.S. degrees in electrical engineering from the University of Texas at Dallas in 2021 and 2017, respectively. He is currently a Research Engineer at Toyota Motor North America, InfoTech Labs. His current research interests are focused on in-vehicle systems and advancements in intelligent vehicle technologies.  
\end{IEEEbiography}
\begin{IEEEbiography}
    [{\includegraphics[width=1in,height=1.25in,clip,keepaspectratio]{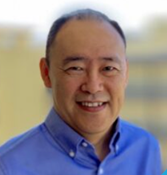}}]
{Kentaro Oguchi} received the M.S. degree in computer science from Nagoya University. He is currently a Director at Toyota Motor North America, InfoTech Labs. Oguchi’s team is responsible for creating intelligent connected vehicle architecture that takes advantage of novel AI technologies to provide real-time services to connected vehicles for smoother and efficient traffic, intelligent dynamic parking navigation, and vehicle guidance to avoid risks from anomalous drivers. His team also creates technologies to form a vehicular cloud using Vehicle-to-Everything technologies. Prior, he worked as a senior researcher at Toyota Central R\&D Labs in Japan.
\end{IEEEbiography}

\begin{IEEEbiography}
    [{\includegraphics[width=1in,height=1.25in,clip,keepaspectratio]{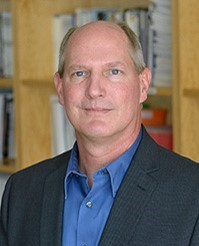}}]
    {Matthew J. Barth}
(Fellow, IEEE) received the M.S. and Ph.D. degree in electrical and computer engineering from the University of California at Santa Barbara, in 1985 and 1990, respectively. He is currently the Yeager Families Professor with the College of Engineering, the University of California at Riverside, USA. He is also serving as the Director of the Center for Environmental Research and Technology. His current research interests include ITS and the environment, transportation/emissions modeling, vehicle activity analysis, advanced navigation techniques, electric vehicle technology, and advanced sensing and control. Dr. Barth has been active in the IEEE Intelligent Transportation System Society for many years, serving as a Senior Editor for both the Transactions of ITS and the Transactions on Intelligent Vehicles. He served as the IEEE ITSS President for 2014 and 2015 and is currently the IEEE ITSS Vice President of Education.
\end{IEEEbiography}

\end{document}